\documentclass[seceq]{ptptex}
\usepackage{graphicx}
\title{%        %You can use \\ for explicit line-break
Theory of the Trojan-Horse Method%
}
\author{%       %Use \scshape  for the family name
Gerhard \textsc{Baur}$^{1}$ and
Stefan \textsc{Typel}$^{2}$%
}
\inst{%         %Affiliation, neglected when [addenda] or [errata]
$^{1}$Institut f\"{u}r Kernphysik, Forschungszentrum J\"{u}lich,
D-52425 J\"{u}lich, Germany\\
$^{2}$Gesellschaft f\"{u}r Schwerionenforschung (GSI),
D-64291 Darmstadt, Germany
}
\abst{%         %this abstract is neglected when [addenda] or [errata]
The Trojan-Horse method is an indirect approach to determine
the energy dependence of S factors of astrophysically relevant
two-body reactions. This is accomplished by studying  closely
related three-body reactions under quasi-free scattering conditions.
The basic theory of the Trojan-Horse method is developed starting from
a post-form distorted wave Born approximation of the T-matrix
element. In the surface approximation the cross section 
of the three-body reaction 
can be related to the S-matrix elements of the two-body reaction.
The essential feature of the Trojan-Horse method is the effective
suppression of the Coulomb barrier at low energies for the
astrophysical reaction leading to finite cross sections at the
threshold of the two-body reaction. In a modified
plane wave approximation the relation 
between the two-body and three-body cross sections 
becomes very transparent. Applications of the Trojan Horse Method are
discussed. It is of special interest that electron screening corrections are 
negligible due to the high projectile energy.
}

\begin{document}

\maketitle

\section{Introduction}

Many astrophysical models depend heavily on precise information about 
nuclear reaction rates that are ideally measured directly in
the laboratory \cite{Rol88}. 
However, cross sections of reactions with charged particles
become very small with decreasing energy due to the Coulomb barrier
and  the astrophysically relevant energy range cannot be reached in 
direct measurements except a few cases. Therefore,
the cross section $\sigma(E)$ at low energies is obtained by extrapolating
experimental data at higher energies with the astrophysical S factor
\begin{equation} \label{Sfac}
 S(E) = \sigma(E) \: E \: \exp(2\pi \eta)
\end{equation}
where $E$ is the c.m.\ energy and
$\eta = Z_{1}Z_{2}e^{2}/(\hbar v)$ is the Sommerfeld parameter
depending on the charge numbers $Z_{1}$, $Z_{2}$ of the colliding nuclei
and their relative velocity $v$.
The extrapolation process introduces uncertainties and important
contributions to the cross section, like resonances, can be missed.
Additionally, a correction has to be applied to obtain the cross section
for bare nuclei because direct laboratory measurements are affected
by electron screening that enhances the measured cross sections
\cite{Ass87,Rol03}.
Independent information on low-energy cross sections is valuable
for a quantitative description of electron screening that is not
yet completely understood.

During the last years, several indirect methods have been developed 
to extract astrophysically relevant cross sections from related 
reactions at higher energies. E.g., the Coulomb dissociation method
\cite{Bau94}
and the method of asymptotic normalization coefficients (ANC)
\cite{Azh01,Xu94} allow to
extract information on low-energy radiative capture reactions.
For general nuclear reactions the Trojan-Horse method (THM) can be applied.
In this approach the astrophysical two-body reaction is replaced by a
suitably chosen three-body reaction that is measured under special kinematical
conditions. The relation between the cross sections is established
with the help of reaction theory.
Without doubt, the indirect process will introduce some uncertainties,
but valuable information can be obtained  on the astrophysical reaction.
Additionally, the errors are independent from that of the direct
measurement. Of course, firm conclusions can be drawn from indirect 
experiments only if the methods have been validated by studying well-known
reactions and if the theoretical approximations are understood
\cite{Aus02}.

%In recent years several indirect methods have been developed 
%to extract cross sections
%relevant to astrophysics  from other types of
%experiments. In these alternative approaches
%the astrophysical relevant two-body reaction is generally 
%replaced by a suitably chosen
%three-body reaction. The relation between the reactions 
%is established with the
%help of nuclear reaction theories. Without doubt, this process will 
%introduce some uncertainties,
%but valuable information on the astrophysical
%reaction can be obtained. Also, the errors are independent 
%from that of the direct approach.
%Of course, the indirect methods have to be validated by studying well known
%reactions before firm conclusions can be drawn from indirect experiments
%in cases where direct measurements are not feasible; see also
%the minireview \cite{Aus02}.

A similarity between cross sections for two-body and closely
related three-body reactions under certain kinematical conditions
\cite{Fuc71}
led to the introduction of the Trojan-Horse method 
\cite{Bau86,Typ00,Typ03}, see also \cite{Bau84,Bau76}.
In this indirect approach a two-body reaction
\begin{equation} \label{APreac}
 A + x \to C + c
\end{equation}
that is relevant to nuclear astrophysics is replaced by a reaction
\begin{equation} \label{THreac}
 A + a \to C + c + b
\end{equation}
with three particles in the final states 
assuming that the Trojan Horse
$a$ is composed predominantly of clusters $x$ and $b$, i.e.\  $a=(x+b)$. 
This reaction can be considered as a special case of a transfer 
reaction to the continuum.
%The observation of a similarity between cross sections for two-body and
%closely related three-body reactions under certain kinematical conditions
%\cite{Fuc71}
%das parallelism bild waere einfach sehr ahnschaulich!
%probleme:(i) einscannen  (ii) copyright
%led to the introduction of the Trojan-Horse 
%method (THM) \cite{Bau86}, see also \cite{Bau84,Bau76}.
%The aim of the THM is to extract the cross section
%of an astrophysically relevant two-body reaction
%\begin{equation} \label{APreac}
% A + x \to C + c
%\end{equation}
%from a suitably chosen reaction
%\begin{equation} \label{THreac}
% A + a \to C + c + b
%\end{equation}
%with three particles in the final state assuming that the Trojan Horse
%$a$ is composed predominantly of clusters $x$ and $b$ : $a=(b+x)$ . 
The energy in the entrance channel of reaction (\ref{THreac}) is chosen
around or above the Coulomb barrier and effects from electron screening
are negligible. Nevertheless, under quasifree kinematical conditions
very small energies can be reached in reaction (\ref{APreac}).
The essential feature of the THM is the suppression of the Coulomb barrier
in the two-body reaction. The cross section of the three-body reaction
remains finite when the c.m.\ energy in the $A+x$ system approaches zero.

%The essential feature of the THM is the actual suppression of the Coulomb
%barrier in the cross section of the
%two-body reaction. The cross section of the three-body reaction
%is not reduced when the c.m.\ energy in the $A+x$ system 
%approaches zero as in 
%reaction (\ref{APreac}). The energy in the entrance channel of (\ref{THreac})
%can be around or above the Coulomb barrier and effects 
%from electron screening 
%are negligible. Nevertheless, very small energies in the  
%reaction (\ref{APreac})
%can be reached.

In section \ref{RT} some general aspects in the theoretical description
of transfer reactions into the continuum are discussed.
This leads to the formulation of the THM theory. In a modified plane-wave
approximation the relation between the cross section of reactions
(\ref{APreac}) and (\ref{THreac}) becomes very transparent.
For details we refer to Ref.~\cite{Typ03}.
Applications of the THM are discussed in section \ref{Sappl}
where also a summary and an outlook are presented.

%In Section \ref{RT} the reaction theory is formulated 
%in the post-form DWBA approach and the relation
%of the T-matrix element of the three-body reaction 
%with the S-matrix elements of the two-body reaction is found.
%In connection with a plane wave approximation, fundamental
%TH integrals appear that are discussed in Section \ref{THI}.
%For technical details concerning the calculation of these integrals
%we refer to  \ref{AppA} and \ref{AppB}.
%The modified plane wave approach is discussed in Section ...
%Applications of the THM are discussed in Section \ref{Sappl}.
%A summary and an outlook are presented in the last Section.

\section{Theory}
\label{RT}

\subsection{Transfer reactions into the continuum in  post-form DWBA}
We assume a three-body model where the target nucleus $A$ interacts
with a projectile $a=b+x$. 
The T-matrix element for the elastic breakup reaction 
\begin{equation}
   A+a \rightarrow A+x+b
\end{equation}
is given 
in the post-form of the distorted-wave Born approximation
(DWBA) as (see also eq.~10 of Ref.~\cite{Bau84})
\begin{equation} \label{eq:Tmat}
 T = \langle \chi_{Bb}^{(-)}(\vec{k}_{Bb})
 \Psi_{B}^{(-)}(\vec{k}_{Ax}) \Phi_{b} | V_{xb} |
 \chi_{Aa}^{(+)}(\vec{k}_{Aa}) \Phi_{A} \Phi_{a} \rangle
\end{equation}
%\begin{eqnarray} \label{eq:Tmat}
%  \lefteqn{T_{\vec{q}_{a} \rightarrow \vec{q}_{b} \vec{q}_{x}} 
%   =} \\ \nonumber & & 
%  \int \!\!\! \int
%  d^{3}r_{bx} \: d^{3}R_{Aa} \: \chi^{(-)\ast}_{\vec{q}_{b}}
%  (\vec{R}_{b-Ax}) \chi^{(-)\ast}_{\vec{q}_{x}} (\vec{r}_{Ax})
%  V_{bx}(\vec{r}_{bx}) \phi_{a}(\vec{r}_{bx})\chi^{(+)}_{\vec{q}_{a}}
%  (\vec{R}_{Aa}),
%\end{eqnarray}
where $B$ denotes the system $A+x$ in the final state. 
$\Phi_{a}$, $\Phi_{b}$, and $\Phi_{A}$ are the bound-state wave functions
of $a$, $b$ and $A$, respectively, and $V_{bx}$ is the potential between
$x$ and $b$.
The $\chi$'s are the scattering wave functions
generated by the appropriate optical potentials. 
This expression for the T-matrix element is quite difficult 
to evaluate in general. At high beam energies 
eikonal methods \cite{Hus85} can be
used to simplify it. For an intermediate model see, 
e.g., Ref.~\cite{Bon99}. It contains
some simple limits, like the Serber model: see, e.g., Ref.~\cite{Bau84}. 
%Of course,
In the distorted waves of eq.~(\ref{eq:Tmat}) the interaction of the target 
with the ``participant'' $x$
as well as the ``spectator'' $b$ is included to all orders in general.

\begin{figure}[b]
\vspace{5mm}
\begin{center}
\includegraphics[width=80mm]{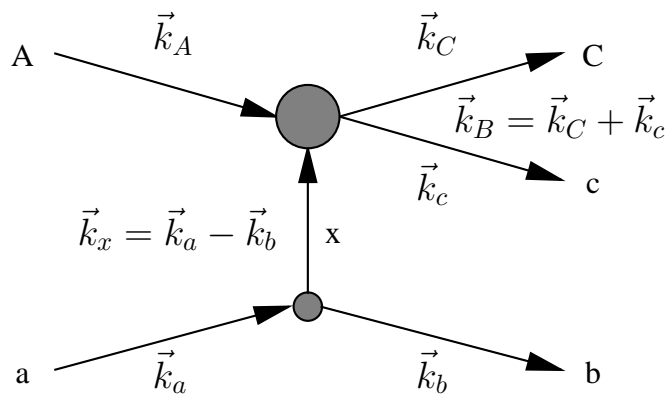}
\end{center}
\unitlength1mm
\begin{picture}(0,0)
\put(50,5){\mbox{\Large $\vec{k}_{a}$}}
\put(85,5){\mbox{\Large $\vec{k}_{b}$}}
\put(40,24){\mbox{\Large $\vec{k}_{x}=\vec{k}_{a}-\vec{k}_{b}$}}
\put(50,50){\mbox{\Large $\vec{k}_{A}$}}
\put(85,50){\mbox{\Large $\vec{k}_{C}$}}
\put(85,30){\mbox{\Large $\vec{k}_{c}$}}
\put(90,39){\mbox{\Large $\vec{k}_{B}=\vec{k}_{C}+\vec{k}_{c}$}}
\end{picture}
\caption{Momenta of the nuclei in the Trojan-Horse reaction (\ref{THreac}).}
\label{fig1}
\end{figure}

It is of interest to treat also the case where the subsystem $B=A+x$ can 
go to other final channels $C+c$. This reaction is sketched in Fig.~\ref{fig1}
with the relevant momenta of the nuclei.
The theoretical description is especially simple when the 
``surface approximation'' can be applied: due to 
Coulomb repulsion and/or strong absorption  the ``wave function of 
the transferred particle''
\begin{equation} \label{eq:wftp}
 \int d\xi_{A} \: \Psi^{(-)\ast}_{B} \Phi_{A} =
  4 \pi \sum_{lm} i^{l} f_{l}(r_{Ax})
 Y_{lm}(\hat{r}_{Ax}) Y^{\ast}_{lm}(\hat{k}_{Cc})
\end{equation}
%f durch f ersetzt, es ist nicht die reg Cou.fkt.
has only to be known in the nuclear exterior. The integration in 
eq.~(\ref{eq:wftp}) is over the nucleon variables of $A$. 
In this case the overlap integral 
%(sometimes also called the wave function of the transferred particle) 
is given in terms of the S-matrix element of the $C+c \to A+x$ reaction,
which we denote by $S_{l}$, as
\begin{equation} \label{eq:wftp2}
 f_{l}(r_{Ax}) = \delta_{AxCc} j_{l}(k_{Ax}r_{Ax})
  + \frac{1}{2} \sqrt{\frac{m_{Ax}k_{Ax}}{m_{Cc}k_{Cc}}} 
  \left( S_{l}-\delta_{AxCc}\right)
  h^{(+)}_{l}(k_{Ax}r_{Ax}) 
\end{equation}
for $r_{Ax}\geq R$ with a cutoff radius $R$.
Here we assume spinless particles for the sake of simplicity.
For charged particles $x$ the appropriate Coulomb functions have to be used
in place of the Bessel (Hankel) functions $j_{l}$ ($h_{l}^{(+)}$).
The validity of the surface approximation was checked by Kasano and 
Ichimura \cite{Kas82}. It was found to be quite good for the (d,p) reaction at 
$E_{d}=26$~MeV. Inclusive breakup spectra were measured for many 
different systems and compared to theory. Agreement is generally
good \cite{Bau84}. 

The theory of inclusive breakup reactions was substantially generalized 
in a series 
of papers by M.~Ichimura, N.~Austern and C.~M.~Vincent (``IAV''). 
We give two references, from where
the full story can be traced back \cite{Ich88,Ich89} . In this series of  
papers, also
many formal aspects have been deeply elucidated and the relation of post-form
to prior-form DWBA (they give identical results) has also been made very clear.

\subsection{Cross section in modified plane wave approximation and THM}

The appearance of the S-matrix element of the two-body reaction in
(\ref{eq:wftp2}) allows to establish a relation between the
cross sections of reaction (\ref{APreac}) and (\ref{THreac}),
see \cite{Typ03} for details.
Replacing the distorted waves in eq.~(\ref{eq:Tmat}) by plane waves
and applying the surface approximation, the cross section for the
three-body reaction 
\begin{equation} \label{d3s}
 \frac{d^{2}\sigma}{dE_{Cc}d\Omega_{Cc}d\Omega_{Bb}}
 = KF \left| W(\vec{Q}_{Bb})\right|^{2} \frac{d\sigma^{TH}}{d\Omega}
\end{equation}
factorizes into a kinematical factor 
\begin{equation}
 KF = \frac{\mu_{Aa}\mu_{Bb}\mu_{Cc}}{(2\pi)^{5}\hbar^{6}} 
 \frac{k_{Bb}k_{Cc}}{k_{Aa}} 
 \frac{16\pi^{2}}{k_{Ax}Q_{Aa}} \frac{v_{Cc}}{v_{Ax}} \: ,
\end{equation} 
a momentum distribution
$|W|^{2}$ and the so-called TH cross section $d\sigma^{TH}/d\Omega$
(see Ref.~\cite{Typ03} for the definition of reduces masses, momenta etc.).
The momentum amplitude
\begin{equation}
 W(\vec{Q}_{Bb}) = -\left( \varepsilon_{a} + 
\frac{\hbar^{2}Q_{Bb}^{2}}{2\mu_{xb}}\right) 
 \langle \exp(i\vec{Q}_{Bb}\cdot \vec{r}_{xb})
 \Phi_{x} \Phi_{b} | \Phi_{a}\rangle
\end{equation}
is related to the wavefunction of the Trojan
horse $a$ with binding energy $\varepsilon_{a} (>0)$
in momentum space. It depends on the momentum 
\begin{equation}
 \vec{Q}_{Bb} = \vec{k}_{Bb} - \frac{m_{b}}{m_{b}+m_{x}} \vec{k}_{Aa} \: .
\end{equation}
Neglecting the Fermi motion of $b$ inside the Trojan Horse the second term 
is the momentum of the incoming 
spectator b with respect to A, and $-\vec{Q}_{Bb}$ corresponds to 
the momentum transfer to the spectator b. 
%In the case of an infinitely heavy target nucleus $A$ this corresponds
%to the relative momentum $\vec{k}_{xb}$ between
%particle $x$ and the spectator $b$.
The momentum distribution essentially describes the Fermi motion of 
$b$ and $x$ inside the Trojan horse $a$.
The TH cross section
\begin{equation} \label{THcs}
 \frac{d\sigma^{TH}}{d\Omega} = \frac{1}{4k_{Cc}^{2}}
 \left| \sum_{l} (2l+1) P_{l}(\hat{k}_{Cc} \cdot \hat{Q}_{Aa})
 \left[ S_{l} J_{l}^{(+)} - \delta_{AxCc} J_{l}^{(-)}\right]
 \right|^{2}
\end{equation}
with Legendre polynomials $P_{l}$ and the S-matrix elements $S_{l}$
looks very similar to the cross section for the inverse of the
astrophysical reaction (\ref{APreac}) except for the TH integrals
\begin{equation}
 J_{l}^{(\pm)}(R,\eta_{Ax},k_{Ax},Q_{Aa}) 
 = k_{ax} Q_{Aa} \int_{R}^{\infty} dr \: r \: u_{l}^{\pm}(\eta_{Ax}; k_{Ax}r)
 \: j_{l}(Q_{Aa} r)
\end{equation}
with the Coulomb wave functions 
$u_{l}^{\pm} = e^{\mp \sigma_{l}} (G_{l} \pm i F_{l})$. 
The TH integrals 
depend on the cutoff radius $R$ of the surface approximation,
the c.m.\ momentum $k_{Ax}$ in the $A+x$ relative motion and
\begin{equation}
 \vec{Q}_{Aa} = \vec{k}_{Aa} - \frac{m_{A}}{m_{A}+m_{x}} \vec{k}_{Bb} \: .
\end{equation}
that reduces to $-k_{x}$ for target mass $m_{A} \to \infty$.
The properties of the TH integrals are discussed extensively in 
Ref.~\cite{Typ03}.

The expression (\ref{d3s}) resembles the form of the cross section
in a plane-wave impulse approximation \cite{Jai70} that has been used
in the past in order to extract information on the momentum distribution
of nuclei. However, only the DWBA with the surface approximation
explains the effective reduction of the Coulomb barrier for small
c.m.\ energies in the $A+x$ system.

\subsection{Threshold behaviour of cross sections}

The energy dependence of the two-body cross section 
\begin{equation}
 \frac{d\sigma}{d\Omega} = \frac{1}{4k_{Ax}^{2}}
 \left| \sum_{l} (2l+1) P_{l}(\hat{k}_{Cc} \cdot \hat{k}_{Ax})
 S_{l}  \right|^{2}
 \propto k_{Ax}^{-2} \exp(-2 \pi \eta_{Ax})
\end{equation}
for the inelastic reaction (\ref{APreac}) 
is governed by the $k_{Ax}^{-2}$ factor  and the
energy dependence 
$S^{l} \propto \exp(-\pi \eta_{Ax})$
of the relevant S-matrix element. This motivates the introduction
of the astrophysical S factor (\ref{Sfac}) for the extrapolation
of experimental data to low energies.
In the TH cross section (\ref{THcs})
the factor $k_{Ax}^{-2}$
is replaced with $k_{Cc}^{-2}$ and the TH integrals $J_{l}^{(\pm)}$ appear.
Their energy dependence for small $k_{Ax}$ is determined by
$k_{Ax}\sqrt{v_{Ax}}\exp(\pi \eta_{Ax})$ 
from the contribution of the irregular Coulomb wave function.
This leads to a $k_{Ax}$ dependence of the
three-body cross section (\ref{d3s}) according to
\begin{eqnarray} 
 \frac{d^{3}\sigma}{dE_{C}d\Omega_{C}d\Omega_{c}} 
 & \propto & k_{Ax}^{-2} v_{Ax}^{-1} \exp(-2\pi \eta_{Ax})
  k_{Ax}^{2} v_{Ax} \exp(2\pi \eta_{Ax}) = \mbox{const.}
\end{eqnarray}
in the lowest order of $k_{Ax}$.
As a result the cross section does not vanish at the threshold
but takes on a finite value. 
Also in the case of neutron transfer,
like in a (d,p) stripping reaction, it is well known
that the cross section is finite
at the threshold $E_{n}=0$ \cite{Bau84,Bau76}. 
The reason is the same as in the case of charged particles:
the momentum dependence of the S-matrix element is cancelled
by the corresponding Trojan-Horse enhancement factor.

In a similar way, the threshold behaviour can be studied in the elastic
breakup case. In this case we have three contributions,
the pure Coulomb, the nuclear and the interference term, see eqs.\
70-73 of \cite{Typ03}. All three terms show the same threshold behaviour, 
the cross section behaves as $k_{Ax} \exp(-2\pi \eta)$ close to threshold.
In contrast, the Coulomb term dominates
in the direct two-body elastic scattering of the $A+x$-system.
The $d+p \rightarrow p+p+n$ breakup reaction was studied recently in the 
relevant kinematical region in \cite{Pel00}.

\subsection{Kinematical Conditions}
\label{Skincon}

In most experiments so far nuclei with 
a dominant s-wave contribution in their ground state
have been employed as Trojan horses. Then, the
momentum amplitude $W(\vec{Q}_{Bb})$ has a maximum at
zero.  Correspondingly, the equation
$ \vec{Q}_{Bb} = 0 $
defines the so-called quasi-free condition in the three-body phase space
where the cross section for the quasi-free reaction reaches a maximum.
From this condition the corresponding quasi-free c.m.\ energy
\begin{equation} \label{erel}
 E_{Ax}^{qf} = E_{Aa} \left( 1 - \frac{\mu_{Aa}}{\mu_{Bb}}
 \frac{\mu_{bx}^{2}}{m_{x}^{2}} \right) - \varepsilon_{a}
\end{equation}
in the initial channel of the two-body reaction  (\ref{APreac})
is derived  from energy conservation.
The relation between $E_{Ax}^{qf}$ and $E_{Aa}$ is purely 
a kinematical consequence. It is obvious that even with a large 
c.m.\ energy $E_{Aa}$
in the entrance channel of the three-body reaction (\ref{THreac})
a small energy $E_{Ax}$ can be reached.
The width of the momentum amplitude
$W(\vec{Q}_{Bb})$ determines the range of energies around
$E_{Ax}^{qf}$ that can be explored due to the Fermi motion of
$b$ and $x$ inside the Trojan horse $a$.
In an actual experiment a cutoff in the momentum
$\vec{Q}_{Bb}$ is chosen to select the region where the
quasi-free process dominates the cross section over all processes.

%\subsection{Elastic scattering with the Trojan-Horse method}
%The main aim for applying the TH method is the extraction
%of the energy dependence of cross sections or astrophysical S factors
%for inelastic two-body reactions. But also the indirect investigation
%of elasting two-body scattering $A+x \to A+x$ can be rewarding. In the direct 
%two-body scattering process the cross section is dominated by the contribution
%of the Coulomb scattering amplitude (\ref{fC}) at low energies that diverges
%with $k_{Ax}^{-2}$. By way of contrast, the TH Coulomb scattering amplitude
%(\ref{fCTH}) vanishes for $k_{Ax} \to 0$ due to the appearance
%of the TH integrals with the regular Coulomb wave function $F_{l}$.
%The nuclear contribution to the TH cross section (\ref{THxs2})
%becomes dominant because of the TH integrals with the
%irregular Coulomb wave functions in eq.\ (\ref{xsTHN}). 
%This allows the study of
%nuclear effects in the scattering at small energies, e.g.\ the influence of
%sub-threshold or low energy resonances. 
%First attempt in this direction were made
%in recent experiments \cite{Spi00,Pel00}.

\section{Applications of the Trojan-Horse Method, Summary and Outlook}
\label{Sappl}

Several reactions have been studied with the TH method recently
\cite{Mus01,Spi01,Lat01,Ali00,Cal97,Spi00,Che96,Spi99,Pel00}.
with ${}^{2}$H and ${}^{6}$Li ($=\alpha +$d)
as typical ``Trojan Horses''.
These nuclei allow to study the transfer of
protons, neutrons, deuterons and $\alpha$-particles, which covers most
of the cases of astrophysical interest for the two-body reaction.

In nuclear astrophysics, transfer reactions (like (d,p) or (${}^{3}$He,d),
or (Li,$\alpha$)) are used  to study resonant states.
E.g., in the ${}^{22}$Na(${}^{3}$He,d)${}^{23}$Mg reaction 
states near the proton threshold were studied \cite{Sch95}. This is relevant 
for the hydrogen burning of ${}^{22}$Na.
In principle, also 
the continuum can be studied. E.g., the ``parallelism'' of (d,p) and 
(n,n) reactions 
has been beautifully shown already in 1971, see Ref.~\cite{Fuc71}. 
The d+${}^{6}$Li reaction was investigated in Ref.~\cite{Che96} 
in this indirect way. Another recent
application is given in Ref.~\cite{Spi99} to the 
${}^{7}$Li(p,$\alpha$)${}^{4}$He-reaction.
An especially interesting case would be the indirect
study of the ${}^{12}$C($\alpha,\gamma$)${}^{16}$O 
reaction by means of a (${}^{7}$Li,t) or (${}^{6}$Li,d) reaction.
Quite recently \cite{Bru01}
the sub-Coulomb  $\alpha$-transfer reaction (${}^{6}$Li,d) and 
(${}^{7}$Li,t) to the bound
$2^{+}$ and $1^{-}$ states in ${}^{16}$O has been used to obtain 
information on the astrophysical S-factor.

In this contribution, the basic theory of the Trojan-Horse method was
reviewed starting from a distorted wave Born approximation
of the T-matrix element. The essential surface approximation
allows to find the relation between the cross section of the
three-body reaction and the S-matrix elements of the
astrophysically relevant two-body reaction. In the 
modified plane wave approximation
the relation between the three-body and two-body cross sections
becomes very transparent. The three-body cross section is
a product of a kinematical factor, a momentum distribution and
a so-called ``Trojan-Horse'' two-body cross section. 
The energy dependence of the appearing Trojan-Horse integrals
leads to a finite cross section of the
three-body reaction at the threshold of the two-body reaction
without the suppression by the Coulomb barrier. This allows to
extract the energy dependence of astrophysical cross sections
from the three-body breakup reaction to very low energies
without the problems of electron screening and extremely low
cross section. A comparison of results for S factors from direct and
indirect experiments can improve the information on the electron 
screening effect, see also \cite{autu}. However, dedicated Trojan-Horse experiments 
are necessary in order
to achieve a precision comparable to direct measurements.

The validity of the Trojan-Horse method can be tested by
comparing the cross sections extracted from the indirect
experiment with results from direct measurements of well studied
reactions.
In principle it is possible to assess systematic uncertainties 
of the Trojan-Horse method by studying various combinations
of projectile energies, spectators in the Trojan Horse
and scattering angles.
Furthermore, different theoretical approximations can be compared,
e.g. full DWBA calculations with and without the surface approximation
and simpler modified plane wave approximations.

One may also envisage applications of the Trojan-Horse method
to exotic nuclear beams. 
An unstable projectile hits a Trojan-Horse target allowing to
study specific reactions on exotic nuclei.
We mention the d($^{56}$Ni,p)$^{57}$Ni 
reaction studied in inverse kinematics in 
Ref.~\cite{Reh98} . In this paper stripping to bound states was studied;
extension to stripping into the continuum would be of interest for this and
other reactions of this type.

A study of low-energy elastic scattering with the Trojan-Horse
method opens another application which can lead to improved
information relevant to the theoretical description of
nuclear reactions at low energies.

\section*{Acknowledgements}

We gratefully acknowledge the hospitality of Claudio
Spitaleri and his group in Catania where many of the
ideas were shaped. 

%\appendix
%\section{First Appendix} %Empty argument \section{} yields `Appendix'. 
%
%\section{Second Appendix}

\end{document}